# About anomalous g-factor value of Mn related defects in GaAs:Mn

S.M.Yakubenya[1]. K.F.Shtel'makh[2,3].

[1]NRC "Kurchatov Institute", Moscow, Russia

[2]St. Petersburg State Polytechnic University, St. Petersburg, Russia

[3]Ioffe Institute, St.Petersburg, Russia

*e-mail:seryak56@mail.ru*

*Abstract*. The results of experimental investigations of **ESR** spectra of manganese impurity ions in a GaAs : Mn system are presented. The studies are done for various a Fermi level position relative to valence band edge in the system. Characteristic defects for the system that give rise to lines with g factors of 5.62 and 2.81 in the **ESR** spectra are studied in some detail. The experimental results are discussed in the framework of a previously developed model with a double defect involving the impurity ion. The "$3d^5$ + hole" model is a special case of the double defect model in this system. An analytical expression for the covalent renormalization of the g factor of an **ESR** line in this system is obtained.

**1. Introduction**

The investigation of $A^3B^5$ semiconductor compounds doped by different impurity ions have been studied for long time and some monograph were published [1,2] . However, prediction of levels position in gap, arising as result of a defects recharge is unsolved problem , for a impurity ions occupied various position in a crystal lattice at the present time.

We formulated a new approach to solving this problem, which has obtained the name "spin marker method"[3,4].

The essence of this method is as follows. Energy levels arising in the forbidden band of a semiconductor as result of the defect recharge can be associated with changes in the charge density in both atom-like orbitals of the defect (an impurity like state) and in orbitals formed predominantly from bands states ( vacancy-like state). In most cases, the second type of state is realized. One exception is impurities with unfilled 3d, 4d, and 4f shells. The charge state of the defects can be identified using resonance methods (**ESR, ENDOR,** etc). Resonance methods allow determination of the total number of electrons localized in a defect for a given position of the Fermi level in the forbidden band , but finding a defect in some charge state depends only on the locations of the Fermi level in the gap and of the defect position in the crystal lattice.

When certain conditions are fulfilled [5,6], there is a "pinning effect" for the position of an energy level in a forbidden band, i.e., the position of the level is virtually independent of the nature of the defect (it does not depend on whether or not the defect has filled d and f shells). The "spin marker method" is based on this similarity of the optical properties of the semiconductor doped by magnetic and non-magnetic ions [7,8].

Far from all magnetic ions with unfilled 3d or 4f shells show the pinning effect; in other words, not all magnetic ions can be used as magnetic "markers" for a given material. For silicon, which was studied in [7,8], the ion chosen as the marker was manganese. In the case of gallium arsenide, considered here, the experimental data are substantially less extensive. Questions that in the cause of silicon already had unambiguous answers before the spin marker method was developed remain to be answered for gallium arsenide and other compositions of $A^3B^5$ before it will be possible to construct a spin marker apparatus for these compositions. The most interesting composition from our point of view is gallium arsenide with hole- type conductivity, with the Fermi level located near the top of the valence band. The first question we must elucidate in this case is in which charge states and for which positions of the ion in the gallium arsenide crystal lattice can be searched out impurity ion if the Fermi level approximately at the top of the valence band. We again choose manganese as a "magnetic marker". We shall take the next step on a way to this purpose in the paper.

Now, the interest to the given system is caused by a possibility of such materials using for a spintronic applications also [9].

Studies of the electronic structure of defects arising in gallium arsenide doped by manganese have continued for more than fifty years [8-19]. It is necessary to note works seeking to investigating of a time depended processes in the system, separately [18,19]. However, detailed analysis of the results received is possible in that case only when one-to-one correspondence between many-electron state of the magnetic defect and corresponding ESR spectrum, regularly observable in experiment will be established, at our opinion.

Otherwise it reminds "to take shots in the dark " more, than that or another. The situation is a little bit better only for the substitution defects with electron configuration $3d^5$, but the defects is not shall be discussed here.

One of the purposes of the paper is finding of one-to-one correspondence between many-electron state of the magnetic defect and corresponding ESR spectrum. The nature of such transitions (electric dipole transition or the magnetic dipole transition) shall be discussed too.

Unfortunately, we do not know original articles, executed for last 20 years, are devoted to multi electrons configuration of manganese ions investigation in various charge state and located in different positions in the lattice.

Even the earliest works on this topic revealed a number of properties distinguishing this impurity from other iron-group transition element impurities in gallium arsenide, in particular, the dependence of the thermic activation energy of the acceptor state (associated with recharge of the manganese impurity ions) on the concentration of the background donors [10]. Blekemore *et al* [10] related this effect to the substitution center of the $Mn_s^k$ (further, $Mn_s$, where s denotes substitution, k –charge state with respect to lattice ), and identified the corresponding level as an acceptor level, located in the gap, $Mn_s^{0/-}$.

It is confirmed by ESR spectra of the samples GaAs doped by manganese + tellurium also [20]. Tellurium is shallow donor in the system. g-factor for the defect $Mn_s^-$ is equal to 2. ($3d^5$ is many-electron configuration of the 3d shell of $Mn_s^-$ defect).

In the same time, the structure of the wave function of the one-electron states that are filled during recharge of the such defect remains unknown. In other words, it is not known whether recharge processes involving the $Mn_s$ defect are accompanied by changes in the degree of filling of the 3d shell of the impurity ion (an impurity-like state) or if they predominantly redistribute the electron density of the bands (a vacancy-like state). There are accordingly two models for the $Mn_s^0$ defect in the literature: "$3d^5 + h_v^+$" [12,14] and "$3d^4$" [21]. Small hybridization between impurity and band states is a basic postulate of the "$3d^5 + h_v^+$" model, but vice versa for the "$3d^4$" one.

We consider a more general double-defect model. . Hybridization parameter is a free adjustable parameter and different positions of the impurity ions in a crystal lattice are considered here. The results of experimental studies of **ESR** spectra of GaAs : Mn samples ( signal with g-factor 2 , signals with g-factor 5.62 and 2.81) are discussed in the framework of this model.

Whether most of the manganese impurity ( located in the different position of crystal lattice ) is in the neutral or ionized state depends on the conditions for doping the original crystal (additional doping together with manganese from shallow acceptor or donor impurities for the example). We will discuss the **ESR** lines with g factors 5.62 and 2.81 in the most detail; most researchers associate these lines with transitions involving terms with $J = 1$ (J is the total angular momentum). In the frame work of the double defect model, there can be two types of defect involving manganese ions for which a $J = 1$ term is the ground state. In addition, we consider the influence of hybridization between impurity and band states on the parameters of the ESR spectra of defects.

## 2. Model for the manganese impurity centers

We will analyze our paramagnetic resonance data using the double defect model proposed for a Si : Mn system[7,8]. Two different defect types in semiconductors irrelevant with each to other are discussed as a usual [2]. It is named by interstitial defects -{$X_i$} (impurity ions located in a crystal lattice pore) and substitution defects {$X_s$} (impurity ions located in point of lattice). In the same time, in the framework of the double defect model, the interstitial and substitution defects are considered as a two limiting cases of double defect. A gallium vacancy {$V_{Ga}$} acts as a partner for the Mn impurity ions {Mn$_s$} in the suggested double defect. At that when,

--- distance between partners is infinity, then interstitial defect type is realized;
--- distance between partners is an infinitesimal value, then we have substitution type of defect.

The account of gallium vacancy at interstitial defect case is necessary to consider all type of double defects from the common base. This will result to unique parameter - Fermi level position ($E_F$), which will define a degree of one-electronic orbital fillings of such kind of double defect. ($E_F$) is counted from vacuum level position in the case.
The process of substation defect formation within the framework of double defects differ from those within framework of the standard model. It is formed as though in two stages - from unit the ion of a native lattice and impurity ion recombine with the formed vacancy already.

As result, numbers of electrons, located on valence bonds between the central ion and the nearest neighbors, ($n^v$) may be differ from one to one for the impurity ion cases and ions of the crystal lattice. Dangling bonds of the nearest ligands (DHH –dangling host hybrids) push out from the valence band in a gap in the case.

We will use notation DHH (dangling host hybrids) by analogy with DBH (dangling bond hybrids) which is applied for the description of the located states in a gap , arising at interstitial defects recharge [22] and VBH (valence bond hybrids) for states of valence band .

The parameter ($n^v$) is defined by binding energy of an electrons in free ion X. Impurity ion electrons with binding energy more than $E_{cr}$ (critical energy) can't valence bond form because its anti-bonding orbitals will be located above than vacuum level . $E_{cr}$ is smaller or equal to third ionization potential of Ga for a cation sublattice in GaAs compound.

The intermediate case, when distance between partners is of the order of the lattice period, the only one valence bond saturated exists between partners ($n^v$ =1). Such type of defects has been named by the pairing defects. A plenty of pairing defect different types may exist, depending on overlapping integral of wave functions partners and number of nonequivalent crystallographic positions in a lattice. Only single type of pairing defects is realized in GaAs. Moreover, we can see electronic density redistribution between the central ion and ions of the nearest crystal environment as a result of temperature variation or other factors, in particular, a electronic configuration change of everyone components as result of Jahn-Teller distortions in the frame work of the model .It is necessary to note, that electronic density redistribution can be realized and without a free carrier generation .The magnetic-moment average value located at manganese ion can be renormalized very strongly as free carriers appear in the system

The considering of double defect model we intentionally limit to the consideration of single elementary cell, understanding thus, that significant electronic density delocalization take place due to hybridization effects and indemnification of a charge occurs much further, than the first coordination sphere. Here and further for a designation of such defects we shall use the notation for substitution {Mn$_s^m$ - V$_{Ga}^n$} and { Mn$_i^m$ - V$_{Ga}^n$ } for pairing defects , {Mn$_i^m$} for interstitial defects . **m** and **n** indexes designate a charge of everyone components with respect to its nucleus.

As a result of the interaction of the 3d states with the crystal field and their hybridization with the orbitals of ligands, crystal field resonance **(CFR)** and dangling host hybrid **(DHH)** states form [2,22]. In the case of manganese ions ,substituted site of a cation sublattice , this primarily has relevance for the one-electron states ($t_2^d$) which form when the 3d orbital are split by the crystal field of $T_d$ symmetry, and ($t_2^v$), which are vacancy-like states whose wave function is made up of p state bands. The states $(t_2)^{VBH}$ and $(t_2)^{DHH}$ have the structure [22]:

$$|t_2^{DHH}\rangle = \zeta |t_2^d\rangle - \eta |t_2^v\rangle$$

(1)

$$|t_2^{VBH}\rangle = \eta |t_2^d\rangle + \zeta |t_2^v\rangle$$

here η and ζ are constants determining the contribution of each component to the hybridized wave function. In the simplest case, when the integral overlap of η and ζ is small, we can obtain the analytical expressions [1]:

$$\zeta^2 = \frac{1}{2}\left[1 + \frac{\delta}{(\delta^2 + V_0^2)^{1/2}}\right] \quad (2)$$

$$\eta^2 = \frac{1}{2}\left[1 - \frac{\delta}{(\delta^2 + V_0^2)^{1/2}}\right]$$

with the normalization condition

$$\zeta^2 + \eta^2 = 1 \quad (3)$$

where $\delta$ is the energy interval between the one-electron orbital ($t_2^d$) and ($t_2^v$) without taking account of hybridization and $V_0$ is a hybridization parameter.

Relations similar (1) ÷ (3) can be written for bonding (VBH) and anti-bonding orbitals (CFR). DHH or CFR orbitals are pushed out in the gap in depend on a relation between η and ζ.

a) **Substitution defects**

Taking this into account and using the notation for the ($t_2^{CFR}$), ($t_2^{DHH}$) and ($e_2^{CFR}$) orbital introduced above (the last states, which are doubly degenerate in the quantum number L, form due to the splitting of the 3d orbital of the crystal field of $T_d$ symmetry), we can write the electronic configuration for the neutral (with respect to the lattice) substitution defect [$Mn_s^0 - V_{Ga}^0$] in the form { $(e_{2+}^2 t_{2+}^2)^{CFR} - (a_1^2 t_2^6)^{VBH}$ } in the manganese center with $3d^4$ electronic configuration model [20], or { $(e_{2+}^2 t_{2+}^3)^{CFR} - (a_1^2 t_2^4)^{VBH} (t_{2-}^1)^{DHH}$ } in the "$3d^5 + h_v^+$" model [12, 14]. Here and below, the notation $\varphi^{VBH}$ ($\varphi = a_1, t_2$), as it was told above already, is used for $sp^3$ bond orbital (the crystal valence state), and $\psi_{2+}^k$, $\psi_{2-}^k$ ($\psi$ = e, t) is used for orbitals with "up" and "down" spin and of an occupation degree by electrons k.

It is interesting that in the case of a [$Mn_s^0 - V_{Ga}^0$] defect with electronic configuration { $(e_{2+}^2 t_{2+}^3)^{CFR} - (a_1^2 t_2^4)^{VBH} (t_{2-}^1)^{DHH}$ } the ground state J=1 term is a hybrid of the ($t_2^d$) and ($t_2^v$) states. Thus, there is no need to make additional assumptions about the strength and sign of the exchange interaction between the 3d core and the band electrons (see, for example, [12, 14]). In other words, this term can be the ground state of a [$Mn_s^0 - V_{Ga}^0$] defect in the **1-s** bond approximation, as for other 3d impurities in gallium arsenide, and not in the **j-j** bond approximation, as proposed in the "$3d^5 + h_v^+$" model [12, 14].

When this term is the ground state of a [$Mn_s^0 - V_{Ga}^0$] defect (electronic configuration { $(e_{2+}^2 t_{2+}^3)^{CFR} - (a_1^2 t_2^4)^{VBH} (t_{2-}^1)^{DHH}$ }), the J = 2 term will be the nearest excited state, and its electronic configuration will have the form { $(e_{2+}^2 t_{2+}^2)^{CFR} - (a_1^2 t_2^4)^{VBH} (t_{2-}^1 t_{2+}^1)^{DHH}$ }. Here we used notation $(t_{2+}^1 t_{2-}^1)^{DHH}$ i.e. the magnetic moment of electrons, occupied such orbital, is equal to zero, since for p electrons the energy of Coulomb collisions of electrons with parallel spins is greater than the energy of the exchange interaction (a low-spin state). This indicates that the spin of an electron occupied state $(t_{2-}^1)^{DHH}$ (the ground state of a [$Mn_s^0 - V_{Ga}^0$]) defect with electron configuration { $(e_{2+}^2 t_{2+}^3)^{CFR} - (a_1^2 t_2^4)^{VBH} (t_{2-}^1)^{DHH}$ } is directed opposite to the spin of the electrons occupied the $(t_{2+}^3)^{CFR}$ orbital, since the hybridization process occurs without "spin-flip" process.

As a result, full the magnetic moment J of such a defect in the ground state equal to

$$J = S + (-1) \cdot L = (5/2 - 1/2) + (-1) \cdot 1 = 1. \qquad (4)$$

The factor (-1) in front of the angular momentum L is due to the fact that the p and d electrons have different parity [16]. *Alias, ground state of the magnetic ions is low spin state.*

The defect $[Mn_s^- - V_{Ga}^+]$ will be ground state of substation defect if negative –U properties is realized for the gallium vacancy. Electronic configuration of such defect will have the form $\{(e_{2+}^2 t_{2+}^3)^{CFR}(t_{2-}^1)^{DHH} - (a_1^2 t_2^2)^{VBH}\}$ and full magnetic moment J of it will be equal to 1 too.

In the case of a center $[Mn_s^0 - V_{Ga}^0]$ ("$3d^4$") with electronic configuration $\{(e_{2+}^2 t_{2+}^2)^{CFR} - (a_1^2 t_2^6)^{VBH}\}$, the ground state is the multiplet with $J = 3$, without account for the Janh-Teller effect. (See [21] for a discussion of the possible influence of Janh-Teller distortions on the magnetic moment of this ion.)

b) **Interstitial and paring defects**

The multiplet with $J = 1$ is the ground state of the paring defect $[Mn_i^+ - V_{Ga}^-]$. The electronic configuration of this defect can be written as:

$$\{(e_{2+}^2 t_{2+}^3 t_{2-}^1)^{CFR} - (a_1^2 t_2^2)^{VBH} (t_{2+}^1 t_{2-}^1)^{DHH}\}$$

or (5)

$$\{(e_{2+}^2 t_{2+}^3)^{CFR}(t_{2-}^1)^{DHH} - (a_1^2 t_2^2)^{VBH} (t_{2-}^1 t_{2+}^1)^{DHH}\}$$

Depending on the spatial distribution of the electron density in this pairing defect (whether the electron density is localized near the defects or displaced from the lattice sites), the local symmetry of the surrounding crystal can either coincide with the symmetry of the unexcited crystal lattice ($T_d$) or reduce to $D_{2d}$ or $C_{3v}$. The similar situation is realized for a case of purely interstitial defect, but we must used notation $(t_2)^{DBH}$ instead of $(t_2)^{DHH}$ for an electronic configuration of a manganese ion in such defect. The orbital $(t_2)^{DBH}$ and $(t_2)^{DHH}$ having same structure, but differ one from another by Colomb correlation energy U. Thus, we can see that the double defect model allows several types of a manganese related defects for which the multiplet with $J = 1$ is the ground state. We will now turn to a discussion of the experimental results in the framework of this model.

**3**. **The samples and conditions for the experiment**

During our experiments, we used samples grown using the method of Chokhral'ski and doped with manganese in the process of crystal growth. Studies of the temperature dependence of magnetic susceptibility(it is described by Cure low and the Hall constant were already performed using some of the samples from this series [20,21]. One sample doped with manganese and tellurium was chosen from a series of such samples, and has parameters identical to the sample studied in [20](The temperature dependence of magnetic susceptibility is described by Cure low, termical energy activation of the manganese centers is equal to $\approx 110$ meV , if it is calculated from the top of valence band ). Table 1 presents the parameters for the study samples. Our investigations of the **ESR** spectra of these samples, which is briefly described in [20], were conducted both at a fixed temperature of 3.8 K .

Dependence of the intensity of **ESR** lines on impurities concentration    Table1

| № | $N_{total}$ manganese ($cm^{-3}$) Mn | Concentration ($cm^{-3}$) | | Concentration (arb.units) | | Specific electrical resistance (ohm cm) |
|---|---|---|---|---|---|---|
| | | $[Mn_s^- - V_{Ga}^0]$ | $[Mn_s^0 - V_{Ga}^0]$ | g=5.62 | g=2.81 | |
| 1* | $4.4 \times 10^{18}$ | $4.4 \times 10^{18}$ | $2.0 \times 10^{16}$ | $2.4 \times 10^{-3}$ | $1.0 \times 10^{-3}$ | $>10^5$ |
| 2 | $1.1 \times 10^{18}$ | $1.0 \times 10^{18}$ | $1.0 \times 10^{17}$ | $4.6 \times 10^{-4}$ | $2.0 \times 10^{-3}$ | $>10^5$ |
| 3 | $2.0 \times 10^{17}$ | $1.0 \times 10^{16}$ | $1.9 \times 10^{17}$ | — | $3.0 \times 10^{-3}$ | $>10^5$ |
| 4 | $1.5 \times 10^{16}$ | $3.0 \times 10^{15}$ | $1.3 \times 10^{16}$ | — | $8.0 \times 10^{-4}$ | $>10^5$ |

1* The sample has been doped by manganese and tellurium

We investigated angular dependences of the intensity dispersion of the ES**R** lines in a special attachment at a fixed temperature of 3.8 K. The vector of the oscillating magnetic field $H_1$ in the rectangular resonator was also directed along the <110> axis. The vector of the oscillating electric field $E_1$ and external magnetic field H in the resonator was also directed parallel one to another (axis Z). Angle θ between axis Z and <100> is equal to $135^0$ in our experiments (initial point). The typical dimensions of the samples used in the experiments were 2 x 2 x 7 mm.

Together with the well-known spectra with g = 2.0023, A = 54 x $10^{-4}$ $cm^{-1}$, and a = 13 x $10^{-4}$ $cm^{-1}$ ("A" and "a" is a constant superfine, fine interactions accordingly ) of GaAs : Mn **ESR** samples [11,24], which are associated with $[Mn_s^- - V_{Ga}^0]$ defects with electronic configuration $\{(e_{2+}^2 t_{2+}^3)^{CFR} - (a_1^2 t_2^6)^{VBH}\}$ number of other spectral features are observed, whose nature is still under discussion. Here, we will discuss defects responsible for the appearance of **ESR** lines with g factors 2.81 and 5.62 (see also [13] for results of studies in GaAs : Mn samples at temperatures below 10 K). Most researchers who have studied such defects [11, 12] relate the observed signals to permitted ($\Delta J_z = 1$) and forbidden ($\Delta J_z = 2$) magnetic dipole transitions in the multiplet with $J = 1$. (Here and below $J_z$, $J_x$, $J_y$, is the projections of the total angular momentum $J$.) The spin Hamiltonian $H_h$ describing the spectrum of magnetic dipole transitions of this defect can be written, using standard notation [23]:

$$H_h = g\beta J_z H_z + D(3J_z^2 - J(J+1)) + E(J_x^2 - J_y^2) + a^1(S_x^4 + S_y^4 + S_z^4 - S(S+1)(3S^2 - 3S - 1)) \quad (6)$$

where $a^1$ is a fine interaction constant for the defect, β is a Bohr magneton. We added the last component to spin-Hamiltonian $H_h$ to describe interaction of the manganese magnetic moment with the nearest ligands. Spin operators ($S_x^4$, $S_y^4$, $S_z^4$) are used here, because full spin S of the defect is equal to 2 and as a result there are nonzero matrix elements in frame work of multiplet with J=1 [23]. The $g$ factors of the observed spectra are isotropic within the range of experimental error; thus, the condition $g_z \beta H >> D$ is fulfilled; i.e., to zeroth approximation, a given center can be treated as a defect, with the local symmetry of the surrounding crystal close to cubic. At the same time, the authors [25] had been proposed interpreted the observed signals as electric-dipole transitions in multiplet with J=1. The spin Hamiltonian $H_e$ describing the spectrum of electric dipole transition can be written as [26]:

$$H_e = E_1 [b\beta(J_x J_y + J_y J_x)] + T(J_x J_y + J_x J_x) \quad (7)$$

where b and T are constants.

The energy levels structure will be identical in both case, but angle dependence of transitions intensity be differ. In the case of a $[Mn_i^+ - V_{Ga}^-]$ defect with electronic configuration $\{(e_{2+}^2 t_{2+}^3)^{CFR} (t_{2-}^1)^{CFR} - (a_1^2 t_2^2)^{VBH} (t_{2-}^1 t_{2+}^1)^{DHH}\}$ the wave function of a multi-electron state with $J = 1$ can be represented in the form $/L_z, S_z\rangle$ [20,23]:

$$|J_z\rangle = |+1\rangle = \sqrt{\frac{3}{5}}|+1,+2\rangle + \sqrt{\frac{3}{10}}|0,+1\rangle - \sqrt{\frac{1}{10}}|-1,0\rangle$$

$$|J_z\rangle = |0\rangle = \sqrt{\frac{3}{10}}|+1,+1\rangle + \sqrt{\frac{2}{5}}|0,0\rangle - \sqrt{\frac{3}{10}}|-1,-1\rangle \quad (8)$$

$$|J_z\rangle = |-1\rangle = \sqrt{\frac{1}{10}}|+1,0\rangle + \sqrt{\frac{3}{10}}|0,-1\rangle + \sqrt{\frac{3}{5}}|-1,-2\rangle$$

where $J_z = L_z + S_z$. In the case of a $[Mn_s^0 - V_{Ga}^0]$ defect with electronic configuration $\{(e_{2+}^2 t_{2+}^3)^{CFR} - (a_1^2 t_2^4)^{VBH}(t_{2-}^1)^{DHH}\}$ or a $[Mn_i^+ - V_{Ga}^-]$ defect with electronic configuration $\{(e_{2+}^2 t_{2+}^3)^{CFR}(t_{2-}^1)^{DHH} - (a_1^2 t_2^2)^{VBH}(t_{2-}^1 t_{2+}^1)^{DHH}\}$ it can be represented in the form:

$$|J_z\rangle = |+1\rangle = \sqrt{\frac{3}{5}}|-1,+2\rangle + \sqrt{\frac{3}{10}}|0,+1\rangle - \sqrt{\frac{1}{10}}|+1,0\rangle$$

$$|J_z\rangle = |0\rangle = \sqrt{\frac{3}{10}}|-1,+1\rangle + \sqrt{\frac{2}{5}}|0,0\rangle - \sqrt{\frac{3}{10}}|+1,-1\rangle \quad (9)$$

$$|J_z\rangle = |-1\rangle = \sqrt{\frac{1}{10}}|-1,0\rangle + \sqrt{\frac{3}{10}}|0,-1\rangle + \sqrt{\frac{3}{5}}|+1,-2\rangle$$

where $J_z = \alpha L_z + S_z$ (where $\alpha = -1$; see above). ). Since the p and d electrons have different parity, the *g* factors of **ESR** lines for transitions with $\Delta J_z = 1$ will have different values for a $[Mn_i^+ - V_{Ga}^-]$ defects with electronic configurations $\{(e_{2+}^2 t_{2+}^3 t_{2-}^1)^{CFR} - (a_1^2 t_2^2)^{VBH}(t_{2+}^1 t_{2-}^1)^{DHH}\}$ and $\{(e_{2+}^2 t_{2+}^3)^{CFR}(t_{2-}^1)^{DHH} - (a_1^2 t_2^2)^{VBH}(t_{2-}^1 t_{2+}^1)^{DHH}\}$.

The *g* factor tensor is a diagonal matrix if we use the following linear combinations for each of the bases (8) and (9) as the wave functions[23]:

$$|+\rangle = \cos\alpha|+1\rangle + \sin\alpha|-1\rangle$$
$$|0\rangle = |0\rangle \quad (10)$$
$$|-\rangle = \sin\alpha|+1\rangle - \cos\alpha|-1\rangle$$

where $tg2\alpha = E/G$ and $G = g_s\beta H$.

According to the definition of the *g* factor,

$$g_z = \langle\Phi|L_z + g_s S_z|\Phi\rangle \quad (11)$$

where $\Phi$ are the wave functions represented by the basis (10) with appropriate weighting factors (see the definitions of $(t_2^{CFR})$ and $(t_2^{DHH})$ and $g_s$ is the *g* factor for a free electron. After straightforward manipulation, we obtain the expressions for the permitted transitions for defects with electronic configurations

$\{(e_{2+}^2 t_{2+}^3)^{CFR}(t_{2-}^1)^{DHH} - (a_1^2 t_2^2)^{VBH}(t_{2-}^1 t_{2+}^1)^{DHH}\}$ and $\{(e_{2+}^2 t_{2+}^3 t_{2-}^1)^{CFR} - (a_1^2 t_2^2)^{VBH}(t_{2+}^1 t_{2-}^1)^{DHH}\}$ respectively[20]:

$$g_z = \frac{1}{2}(3g_z - \eta^2) = 3.0 - \frac{\eta^2}{2} \quad (12)$$

$$g_z = \frac{1}{2}(3g_z + \zeta^2) = 3.0 + \frac{\zeta^2}{2}$$

Given the experimental value of the g factor, 2.81, we find that the only defects that could give rise to this line in the **ESR** spectrum are the $[Mn_s^0 - V_{Ga}^0]$ defect with electron configuration $\{(e_{2+}^2 t_{2+}^3)^{CFR} - (a_1^2 t_2^4)^{VBH} (t_{2-}^1)^{DHH}\}$ or the $[Mn_i^+ - V_{Ga}^-]$ defect with electron configuration $\{(e_{2+}^2 t_{2+}^3)^{CFR} - (a_1^2 t_2^2)^{VBH} (t_{2-}^1 t_{2+}^1)^{DHH}\}$. However, it is not appropriate to classify this state purely as a state of broken bonds, since a good agreement with the experimental data is reached for $\eta^2 \sim 0.38$, which indicates that the p orbitals make an appreciable contribution to the wave function of the hybrid state.

Note that in the framework of the "$3d^5 + h_v^+$" model, the calculated g factor is 2.75 [23,27]:

$$g = \frac{1}{2}[g_s - g_l] + \frac{S(S+1) - j(j+1)}{2J(J+1)}[g_s + g_l] \quad (13)$$

where $g_l = -1$ (the g factor of a hole); $j = 3/2$ is the magnetic moment of a localized hole; $S = 5/2$ is the magnetic moment localized in the manganese ion impurity; $g_s = 2.0023$ is the g factor of the ion with configuration $3d^5$; $J = 1$ is the total magnetic moment localized in the $[Mn_s^0 - V_{Ga}^0]$ defect. It is not possible to improve the agreement with the experimental data, since there are no variable parameters in this expression. In addition, as can easily be seen from (13), the numerical value 2.75 is obtained for the g factor in the "$3d^5 + h_v^+$" model when correct allowance is made for the hybridization between the impurity and band states for $\eta^2 \cdot \zeta^2 = \frac{1}{2}$. This corresponds to the resonance condition between the $(t_2^d)$ and $(t_2^v)$, seed states and the initial super-deep localization of these states relative to the edge of the valence band; this, in turn, corresponds to the first postulate used to construct the model in question [12].

Angular dependence of the transition intensity ($W_1$) will differ for cases of magnetic dipole and electro dipole transitions, as we already spoke above, when we make rotation of the sample around an axis <110> in the plane [110]. It is described by formula [26]:

$$W_1 = C_1[C_2 + E_1\{(b\beta H_0 + T(2J_z - 1))\cos(\theta)\}^2] \quad (14)$$

for the electric dipole transition in the range of multiplet with J=1. Here and in the sequel, angle $\theta$ calculated from axis <100>. $C_1...C_n$ are constants. Another formula for $W_1$ had been used by authors [25] for the electric dipole transition ("$3d^5 + h_v^+$"):

$$W_1 = C_3[\{\cos(\theta)\}^2\{1 - 3\{\sin(\theta)\}^2\{\cos(2(\theta))\}\}] \quad (15)$$

$W_1$ in the framework of the double defect model is describing by means of ( defect $[Mn_i^+ - V_{Ga}^-]$ ) :

$$W_1 = C_{10} + C_4[1 - 5\{\sin(\theta)\}^2 + 3.75\{\sin(\theta)\}^4] \quad (16)$$

for the magnetic dipole transition case.
It arise out of the interaction between magnetic moment located on a manganese ions and electrons located at the nearest ligands.

Angular dependence of the magnetic dipole transition intensity ($W_1$) in the framework of the "$3d^5 + h_v^+$" model (defect [$Mn_s^0$ - $V_{Ga}^0$] ) is describing by means of [25]:

$$W_1 = C_5 [1 - 0.75\{\sin(2\theta)\}^2] \qquad (17)$$

Comparison with experiment shows that the best possible fit to be received in the assumption, that the given transition is a magnetic dipole and angular dependence of its intensity is described by the formula (16) (Fig.1). Alias, it arise from the [$Mn_i^+$ - $V_{Ga}^-$] with electron configuration $\{(e_{2+}^2 t_{2+}^3)^{CFR}(t_{2-}^1)^{DHH} - (a_1^2 t_2^2)^{VBH}(t_{2-}^1 t_{2+}^1)^{DHH}\}$. At the condition is realized $g_z \beta H >> D, E$.

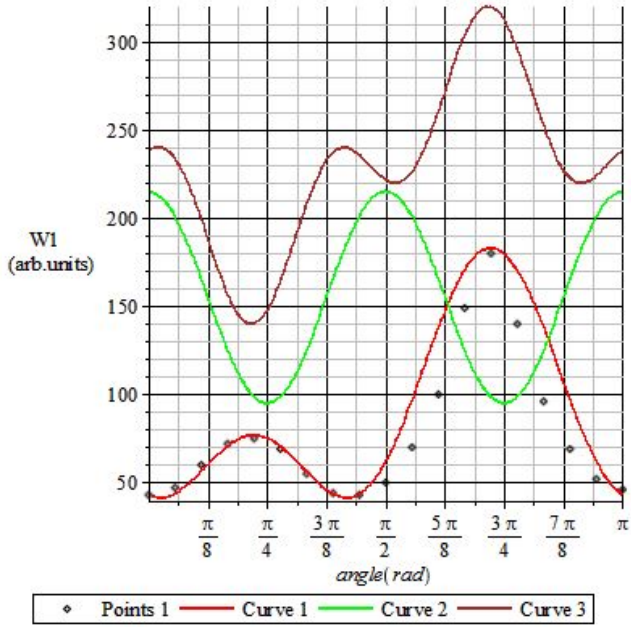

Fig.1
The angular dependence of the transition intensity (W1) of *magnetic dipole and electro dipole transitions* in framework of different models. Point 1 – experimental result (present work) ; Curve 1 calculated – pairing defect, *magnetic dipole transition* $C_{10} + C_4 [1 - 5\{\sin(\theta)\}^2 + 3.75\{\sin(\theta)\}^4]$ (present work) ; Curve 2 calculated --- "$3d^5 + h_v^+$" [25], *magnetic dipole transition* - $C_5 [1 - 0.75\{\sin(2\theta)\}^2]$ ; Curve 3 calculated ---"$3d^5 + h_v^+$" [25], *electric dipole transition* - $C_3 [\{\cos(\theta)\}^2 \{1-3\{\sin(\theta)\}^2\{\cos(2(\theta))\}\}]$ .

The presence of signal with doubled value g factors in the experimental spectra indicates the necessity of including a term in the initial spin Hamiltonian describing the rhombohedra distortion of the local symmetry around the impurity defect, i.e., we must include a term of the form E ($J_x^2 - J_y^2$). In this case, the probability of a magnetic dipole transition $W_2$ with a doubled value of *g* factor ($g_z = 5.62$) is determined by the expression:

$$W_2 = g_z \sin(2\alpha)\{\cos(\theta)\}^2 \qquad (18)$$

$$W_2 = C_6(1+3\{\cos(\theta)\}^2 + 0.75\{\sin(2\theta)\}^2) \qquad (19)$$

in the framework of the double defect model ([$Mn_i^+$ - $V_{Ga}^-$]) [16] and "$3d^5 + h_v^+$" (defect [$Mn_s^0$ - $V_{Ga}^0$] )[25] correspondingly. But, the next expression we will be for the $W_2$ if a transition with $g_z = 5.62$ is arise from the electric dipole:

$$W_2 = C_7 [C_8 + E_1 \{\sin(\theta)\}]^2 \qquad (20)$$

$$W_2 = C_9 \{\sin(\theta)\}^2 [1 + 5\{\cos(\theta)\}^2 + 6\{\cos(\theta)\}^4] \qquad (21)$$

in the frame work of the double defect model ($[Mn_i^+ - V_{Ga}^-]$) [20] and "$3d^5 + h_v^+$" (defect $[Mn_s^0 - V_{Ga}^0]$) [25] correspondingly. Comparison with experiment shows that the best possible fit to be received in the assumption, that *the given transition is a electric dipole and angular dependence of their intensity is described by the formula (20) (Fig.2)*

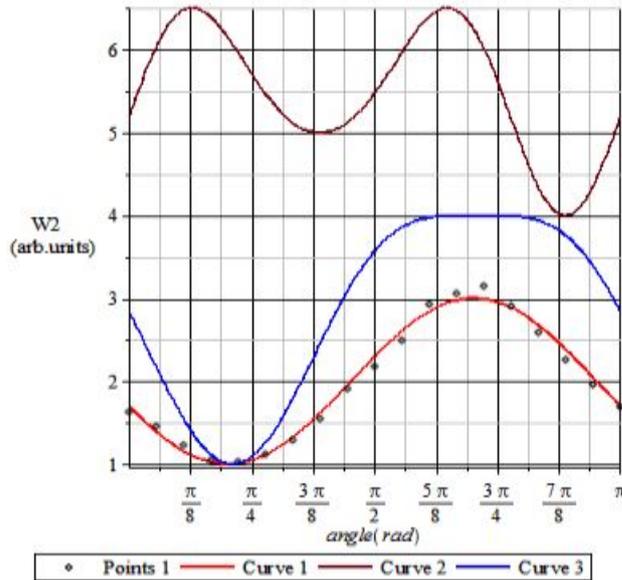

Fig.2
The angular dependence of the transition intensity (W2) of *magnetic dipole and electric dipole transitions* in frame work of different models. Point 1 –experimental result (present work) ; Curve 1 calculated– pairing defect, *electric dipole transition* $C_7[C_8 + E_1\{\sin(\theta)\}]^2$ (present work) ;Curve 2 --calculated -"$3d^5 + h_v^+$" [25] , *electric dipole transition* - $\{\sin(\theta)\}^2[1 + 5\{\cos(\theta)\}^2 + 6\{\cos(\theta)\}^4]$ ; Curve 3-- calculated --- "$3d^5 + h_v^+$" [25] , *magnetic dipole transition* - $C_6(1+3\{\cos(\theta)\}^2 + 0.75\{\sin(2\theta)\}^2)$ ;

Alias, it arise from the $[Mn_i^+ - V_{Ga}^-]$ with electron configuration $(e_{2+}^2 t_{2+}^3)^{CFR} (t_{2-}^1)^{DHH} - (a_1^2 t_2^2)^{VBH} (t_{2-}^1 t_{2+}^1)^{DHH}$ . At once, this transition may be arise from the $[Mn_s^- - V_{Ga}^+]$ if negative U properties is realized for the gallium vacancy. The electron configurations of the defect $[Mn_i^+ - V_{Ga}^-]$ and $[Mn_s^- - V_{Ga}^+]$ is identical almost .

### 4.Conclusion

We have executed investigations of gallium arsenide crystals doped by manganese. The samples had a hole type of conductivity. Several various types of defects were detected in such system. It is shown, that the model of double defects offered for the description of a transition metal ions properties in silicon earlier [7,8] can be successfully applied for the description of their properties in gallium arsenide. We have established, that the model "$3d^5 + h_v^+$" [12,14] are used earlier for the description of such system properties is a special case of model of double defects. It has been found, that the ground state of magnetic ion can be low-spin as result of hybridization between impurity and states. We reached a conclusion , analyzed of angle dependence transition intensity with g-factor 5.62 and 2.81 , that it is electric dipole transition and magnetic dipole transition in range of multiplet with J=1 correspondingly . The given term is the ground state of the defect $[Mn_i^+ - V_{Ga}^-]$ .However, the received results interpretation is not final as ideal concurrence between experimental results and calculations, executed within the framework of the given model it is not achieved, and it is required carrying out of additional researches, taking into account importance of the problem .

**References**

[1]  Lannoo M. Bourgoin.J, Point defect in semiconductor I (Springer-Veriag, Berlin,1981) p.256
[2]  Kikoin K.A. Electronic properties of transition metal impurities in semiconductors,
     ( Energoatomizdat ,Moscow,1991),p.301
[3]  Yakubenya S M 1992 *Fiz. Tverd. Tela* 34  345 (1992)
[4]  Aleksandrov P A and Yakubenya S M *Proc. Int. Conf. Science and Technology of Dejects* Control in
     Semiconductors  *ed.by K Sumino v. 2  (Elsevier Science Publishers B.* V, *Yohogama ,1989*) p 1605
[5]  Yakubenya S M  *Fiz. Tverd. Tela* 34 ,345 (1992)
[6]  Yakubenya S M  *Fiz. Tverd. Tela* 34 , 2753 (1992)
[7]  Yakubenya S M  *Fiz. Tverd. Tela* 33 ,1464 (1991)
[8 ] Yakubenya S M *Fiz. Tverd. Tela* 33  , 1469  (1991)
[9]  Wolos A. , Kaminska M.,Magnetic impurities in Wide Band –gap  III- V Semiconductors
     Spintronics Book  Series: Semiconductors and Semimetals v.82 (edited by Dieti T. ,et.al., Elsevier , 2008)
[10] Brown W J and  Blekemore J S   *J. Appl. Phys.* 43 ,  2242 (1972)    http://dx.doi.org/10.1063/1.1661483
[11] Schneider J, Kaufmann U, Wikening  W, Baeumler M, and Köhl F *Phys. Rev. Lett.* 59 ,240 (1987)
     http://dx.doi.org/10.1103/PhysRevLett.59.240
[12] Masterov V F, Mikhrin S B, Samorukov S B, and Shtel'makh K F *Fiz. Tekh. Poluprov.* 17,1259 (1983)
[13] Almeleh H and Goldstein N  *Phys. Rev.* 128, 1568 (1962)  http://dx.doi.org/10.1103/PhysRev.128.1568
[14] Freyt Th., Maiert M.,  Schneider J., Gehrket M.  J. Phys. C: Solid State Phys. **21** ,5539,(1988)
[15] V.F. Sapega, T. Ruf, M. Cardona   Solid State Communications, **114**, 573( 2000*)*
     doi:10.1016/S0038- 1098(00)00109-5
[16] Fedorych O. M.,  Hankiewicz E. M.,  Wilamowski Z. , Sadowski Z. Phys.Rev.B **66,** 045201 (2002) doi:
     http://dx.doi.org /10.1103/PhysRevB.66.045201
[17] Weiers T. Phys.Rev.B73, 033201,(2006)  doi: http://dx.doi.org /10.1103/PhysRevB.73.033201
[18] Akimov A, . Dzhioev R.I., Korenev V. L. , . Kusrayev Yu. G. , . Zhukov E. A.,  Yakovlev D.R., Bayer M.
     Phys.Rev.**B 80**,081203(R),(2009)    doi: http://dx.doi.org/10.1103/PhysRevB.80.081203
[19] Akimov A, . Dzhioev R.I., Korenev V. L. , . Kusrayev Yu. G. , . Zhukov E. A.,  Yakovlev D.R., Bayer M.
     J. Appl. Phys. **113**, 136501 (2013); http://dx.doi.org/10.1063/1.4795527
[20] Yakubenya S. M.    J.Moscow Phys.Soc. 3,273 (1997)
[21] Andrianov D. G., Bol'sheva Yu. N., Lazareva G. V., Savel'ev A. S., and Yakubenya S. M.
     *Fiz. Tekh. Poluprov.* 17 ,     810 (1983)
[22] Zunger A and Lundefelt U   Phys. Rev. B 27, 1191 (1983)
[23] Abragam A and Bleaney B  Electron Paramagnetic Resonance of Transition Ions
     (Clarendon ,  Oxford, 1970)
[24] Masterov V. F., Mikhrin S. B., Samorukov S. B., and Shtel'makh K. F .   *Fiz. Tekh. Poluprov*.19, 2093 (1985)
[25] Baran N. P., Maksimenko V. M., Semenov Yu. G., Bratus' V. Ya., Markov A. V.   JETP Letters 5,101(1992)
[26] Ludwig G. W. ,Ham F. S.     Phys.Rev. L ett. 8,210 (1962)
[27] Masterov V. F., Shtel'makh K. F., and Barbashov M. N.          *Fiz. Tekh. Poluprov.* 22  654 (1988)

Figure capture

Fig.1 The angular dependence of the transition intensity (W1) of *magnetic dipole and electro dipole transitions* in frame work of different models. Point 1 –experimental result (present work) ; Curve 1 calculated– pairing defect, – pairing defect, *magnetic dipole transition* $C_{10} + C_4 [1 - 5\{\sin(\theta)\}^2 + 3.75\{\sin(\theta)\}^4]$ (present work) ;Curve 2 calculated ---"$3d^5 + h_\nu^+$" [18] , *magnetic dipole transition* - $C_5 [1 - 0.75\{\sin(2\theta)\}^2]$ ; Curve 3 calculated ---"$3d^5 + h_\nu^+$" [18] , *electric dipole transition* - $C_3 [\{\cos(\theta)\}^2 \{1 - 3\{\sin(\theta)\}^2 \{\cos(2\theta)\}\}]$ .

Fig.2 The angular dependence of the transition intensity (W2) of *magnetic dipole and electro dipole transitions* in frame work of different models. Point 1 –experimental result (present work) ; Curve 1 calculated– pairing defect, *electric dipole transition* $C_7 [C_8 + E_1 \{\sin(\theta)\}]^2$ (present work) ;Curve 2 calculated ---"$3d^5 + h_\nu^+$" [18] , *electric dipole transition* - $\{\sin(\theta)\}^2 [1 + 5\{\cos(\theta)\}^2 + 6\{\cos(\theta)\}^4]$ ; Curve 3 calculated ---"$3d^5 + h_\nu^+$" [18] , *magnetic dipole transition* - $C_6 (1 + 3\{\cos(\theta)\}^2 + 0.75\{\sin(2\theta)\}^2)$ ;

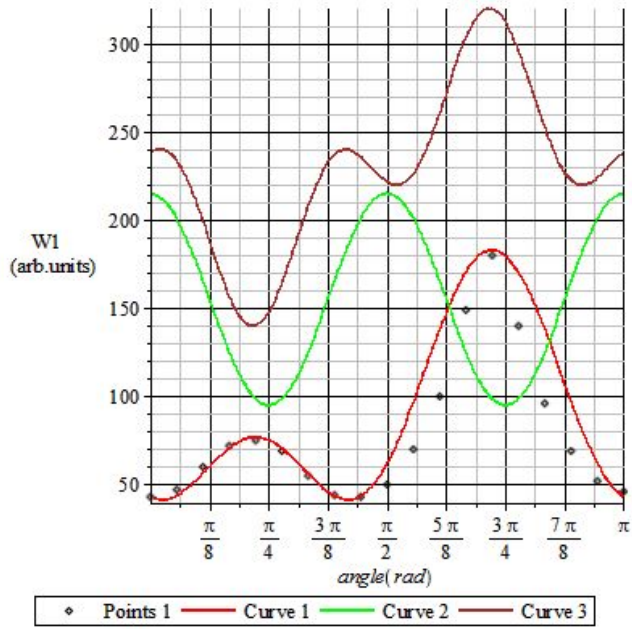

Fig.1

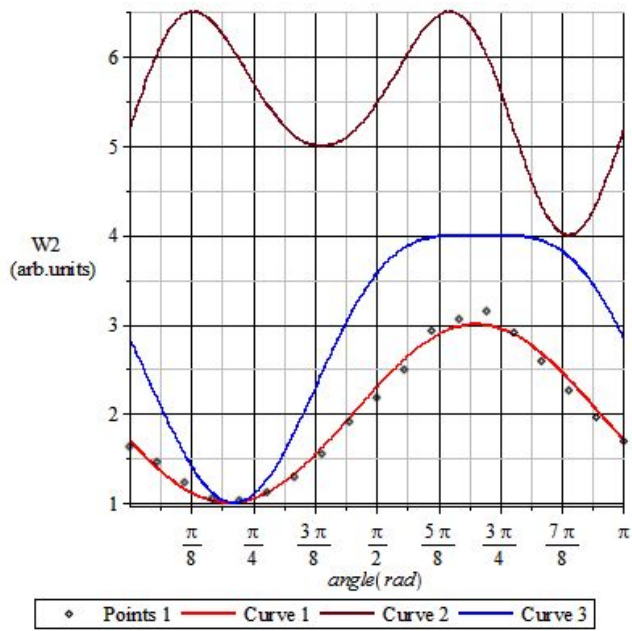

Fig.2